\documentclass[10pt]{iopart}

\usepackage{iopams}
\usepackage[dvips]{graphicx}

\begin{document}

\title[]{Upper limits on stray force noise for LISA}

\author{L Carbone$^a$, A Cavalleri$^b$, R Dolesi$^a$, C D Hoyle$^a$,\\ M Hueller$^a$, S Vitale$^a$ and W J Weber$^a$}

\address{$^a$ Dipartimento di Fisica, Universit\`a di Trento and INFN, Sezione di Padova, Gruppo Collegato
di Trento, I-38050, Povo, Trento, Italy }
\address{$^b$ CEFSA-ITC, I-38050, Povo, Trento, Italy}

\ead{hueller@science.unitn.it, carbone@science.unitn.it}

\begin{abstract}
We have developed a torsion pendulum facility for LISA
gravitational reference sensor ground testing that allows us to
put significant upper limits on residual stray forces exerted by
LISA-like position sensors on a representative test mass and to
characterize specific sources of disturbances for LISA. We present
here the details of the facility, the experimental procedures used
to maximize its sensitivity, and the techniques used to
characterize the pendulum itself that allowed us to reach a torque
sensitivity below 20 fN m $/\sqrt{\textrm{Hz}}$ from 0.3 to 10
mHz. We also discuss the implications of the obtained results for
LISA.

\end{abstract}

\pacs{04.80Nn, 07.10Pz, 07.87+v}

\section{Introduction}
\label{introduction} The LISA (Laser Interferometer Space Antenna)
sensitivity goal requires that the test masses (nominally 2 kg)
are kept in free fall with an acceleration noise below
3$\times$10$^{-15}$ ms$^{-2}/\sqrt{\textrm{Hz}}$ in the frequency
range down to 0.1 mHz \cite{bender}. With most environmental noisy
forces screened by shielding the test masses in a drag-free
satellite, the resulting main source of residual disturbances is
the satellite itself. The spacecraft is kept centered about the
test mass by a displacement sensor which guides a thruster array.
Capacitive position sensors have been developed to meet LISA
requirements in terms of displacement sensitivity and residual
force noise \cite{dolesi,weber:sens}. In parallel with the planned
LTP flight test \cite{LTP:1}, a torsion pendulum facility has been
developed \cite{hueller,carbone} as a precision test-bench for
characterizing, on ground, the purity of free fall allowed by LISA
gravitational reference sensors.

Significant upper limits of the sensor induced force noise can be
placed by suspending a representative hollow test mass inside a
LISA-like position sensor and searching for residual forces
exerted on the test mass along the torsional degree of freedom
down to the level permitted by the pendulum torque noise floor. In
order to maximize the apparatus sensitivity as a torque detector
it is necessary to isolate it from any environmental effect which
could leak into the torsional mode by reducing the coupling to the
disturbance and/or the noise level of the disturbance itself.

In this paper we describe in detail the apparatus, focusing on the
experimental techniques used to characterize and maximize the
sensitivity of the pendulum. As shown in \cite{carbone}, we
reached a torque sensitivity below 20 fN m $/\sqrt{\textrm{Hz}}$
from 0.3 to 10 mHz, within a factor 3 to 5 above the pendulum
intrinsic thermal noise across this frequency range. Finally, we
discuss the implications of these results for LISA displacement
sensor characterization.

\begin{figure}[t]

\begin{center}

\includegraphics[]{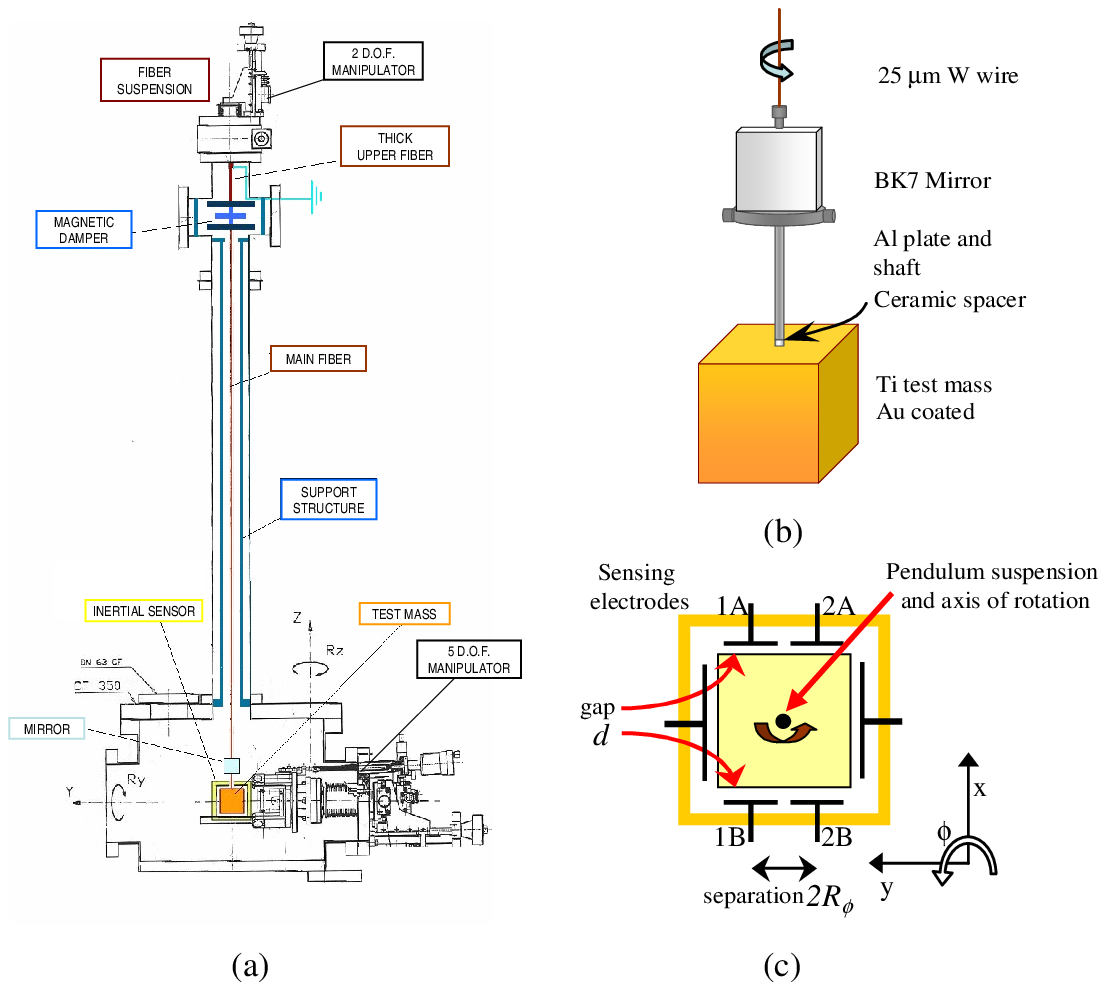}

\caption{(a) Sketch of the experimental apparatus. (b) The test
mass, its support, and the stopper plate that prevents the test
mass from hitting the sensor electrodes. The pendulum has a moment
of inertia $I$ = 338 $\pm$ 5 g cm$^2$ and weighs 101.4 g. (c)
Schematic top view of the sensing electrodes; relevant dimensions
are $R_{\phi}$ = 10.25 mm, and $d$ = 2 mm.}

\label{scheme}

\end{center}

\end{figure}

\section{Details of the torsion pendulum facility}
\label{details}

A schematic of the torsion pendulum facility is shown in
\fref{scheme}a: a vacuum vessel accommodates a prototype sensor
surrounding the test mass and its 6-channel capacitive-inductive
readout electronics \cite{dolesi,weber:sens}. The chamber is
mounted on a platform whose inclination can be adjusted, while the
whole facility sits on a concrete slab partially isolated from
laboratory floor.

As shown in \fref{scheme}b, the torsion pendulum is composed of a
hollow gold-coated Ti cube, with $s$ = 40 mm sides and 2 mm wall
thickness, and a supporting Al bar, on which a stopper plate and
optical mirror for independent readout are mounted. The test mass
is electrically isolated by a ceramic spacer, while the rest of
the pendulum is grounded through the torsion fiber, a Au-coated W
wire nominally 25 $\mu$m thick and 1 m long. The pendulum free
torsional period is $T_0 =$ 515.1 s, with an energy decay time
$\tau_0 \approx 1.35\times10^5$ s, corresponding to a quality
factor Q $\approx$ 1650.

The torsion pendulum hangs from a magnetic eddy current damper
upper stage consisting of a W fiber, with radius $r \approx 50$
$\mu$m and length $l \approx$15 cm, supporting an Al disk
surrounded by toroidal rare earth magnets. The magnetic damper
reduces the swing mode energy decay time to $\approx$ 70 s without
affecting the twist mode performance because of the cylindrical
symmetry of its design. This double suspension acts as a gimble,
ensuring that the main torsion fiber suspension point hangs
essentially vertical, while giving a negligible contribution to
the torsional mode; it is rotationally much stiffer than the main
fiber (the spring constant scales $\Gamma\propto l^{-1} r^4$ for
round fibers).

The capacitive sensor, the Mo-Shapal prototype discussed in
\cite{dolesi}, can be centered, based on the sensor 6-channel
capacitive-inductive readout, around the suspended test mass using
a 5 degree of freedom micromanipulator, while the fiber suspension
point can be raised in $z$ and rotated along $\phi$. The
displacement sensor angular sensitivity, $\approx$ 40 nrad
$/\sqrt{\textrm{Hz}}$, is dominated by intrinsic thermal noise.
The pendulum motion is also monitored by a commercial
autocollimator, with $\approx$ 50 nrad resolution for both twist
and tilt modes, allowing calibration of the sensor by exciting
large twist motion and purposefully tilting the apparatus by a few
$\mu$rad.

\begin{figure}[t]

\begin{center}

\includegraphics[width=12.5cm]{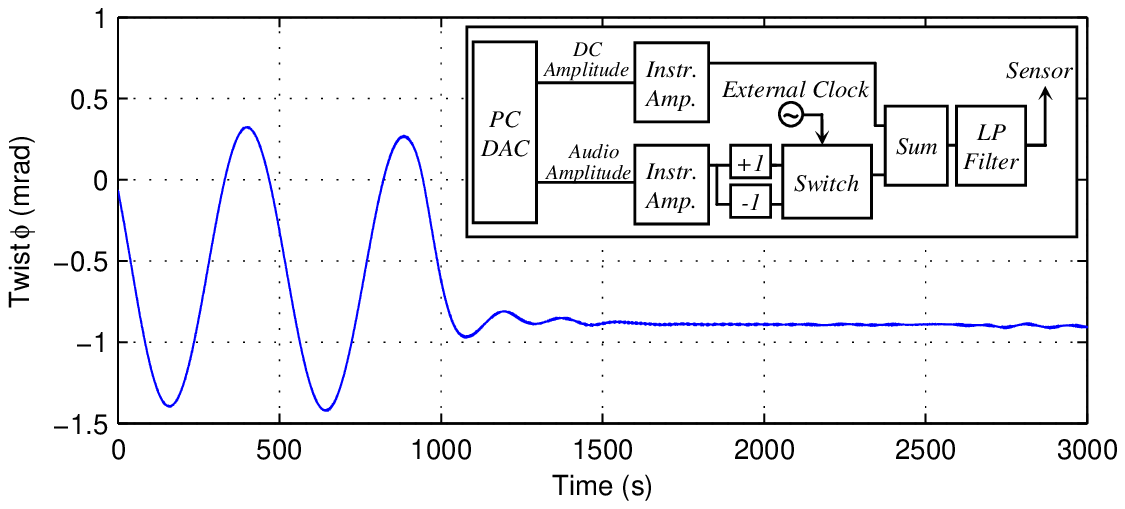}
 \caption{PID control around -0.8 mrad setpoint is applied to the pendulum twist mode
 at t$_0\approx$ 1000 s.
 The inset shows the electrostatic actuation circuitry scheme:
 a switch driven by a 205 Hz external clock alternatively
transmits the amplitude DAC signal or the inverted one, generating
an audio frequency square wave; DC and audio signals are then
summed and low-pass filtered before being applied to the sensor.
To avoid ground loops, the sensor ground is isolated from the DAC
one by a set of instrumentation amplifiers.}

\label{Act-scheme}

\end{center}

\end{figure}

The facility is equipped with home-made electrostatic actuation
circuitry that, as proposed for the LISA actuation scheme
\cite{weber:sens}, is integrated with the sensor bridge
electronics to apply audio frequency and DC voltages to the
sensing electrodes. Audio voltages are used for PID control of the
pendulum torsional mode (see \fref{Act-scheme}), while DC biases
are applied for electrostatic characterization of the sensor
electrodes and to measure the test mass charge
\cite{carbone,weber:DC}. It is worth noting that, as required for
LISA, the electrostatic actuation circuitry does not add excess
noise to the sensor sensitivity.

The pressure is kept below $10^{-5}$ mbar by a vibrationally
isolated turbo pump; use of an ion-pump has been avoided to
prevent electrical charging of the test mass due to electrons
coming from the pump itself. The measured net residual test mass
charging rate, $\approx$ +1 $e$ per second, was occasionally
balanced using electrons emitted by the hot cathode pressure
gauge.

\begin{figure}[t]

\begin{center}

\includegraphics[]{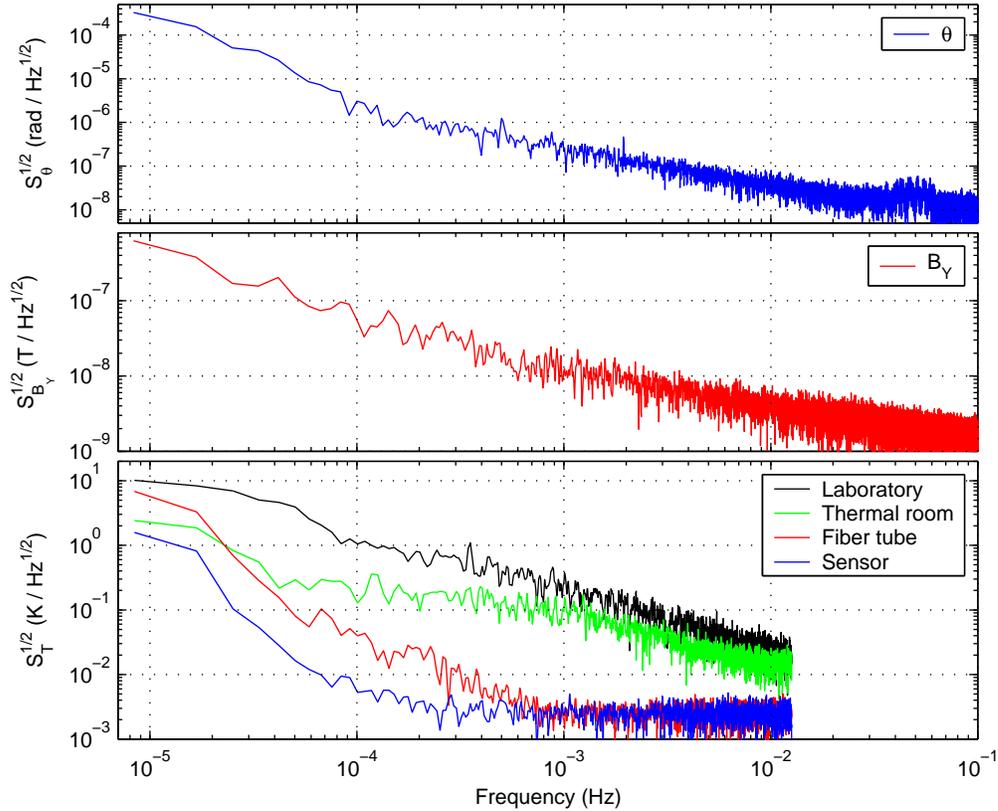}

\caption{Spectral densities of important environmental disturbance
sources. The upper panel shows the fluctuations of lab floor tilt,
as measured by the sensor; tilt noise levels are also
representative of the orthogonal axis $\eta$. The middle panel
shows the fluctuations of one horizontal component of the magnetic
field near the pendulum. The bottom panel shows the temperature
noise at different apparatus locations; the curve labelled
``sensor'' refers to a transducer inside the vacuum, while the
``fiber tube'' thermometer is attached to the outside of the tube.
The temperature data were sampled at 1/40 Hz, whereas all other
sensors where sampled at 10 Hz; the white level at 2.5 mK
$/\sqrt{\textrm{Hz}}$ is the readout noise. The effects of the
stages of thermal insulation are clearly visible.}

\label{envirnoise}

\end{center}

\end{figure}

The entire experiment is enclosed in a thermally insulated room,
whose temperature is controlled by a constant temperature water
bath that stabilizes the air circulating inside a heat exchanger.
The torsion fiber tube is covered with an additional layer of
thermal shielding, giving higher temperature fluctuation
suppression. As shown in \fref{envirnoise}, temperature
fluctuations at the fiber tube are suppressed by at least a factor
20 between 0.1 and a few mHz, and by $\approx$ 10 below 0.1 mHz.
The system is also useful for low temperature bake-out of the
vacuum-chamber (60 $^\circ$C), which reduces the thermally
activated fiber drift from 1 mrad/h to $\sim$ 10 $\mu$rad/h.

The facility is equipped to monitor different environmental
variables. The sensor housing, electronics box, vacuum vessel,
fiber tube, thermal room, and lab temperature are continuously
monitored by Pt100 thermometers. The magnetic field is monitored
by a three-axis 10 nT resolution flux-gate magnetometer, placed in
the neighborhood of the pendulum. The capacitive sensor itself
measures the platform tilt: a tilt of the apparatus along the axis
$\theta$ ($\eta$) will cause a translation of the pendulum
relative to the sensor along $x$ ($y$), determined by the $\sim$ 1
m fiber length, $\Delta x \approx \Delta \theta \times \textrm{1m}
$. Most measurements are automated by dedicated software, and all
experimental data and possible environmental noise sources are
continuously recorded by a (in-house) data acquisition and control
system.

\section{Environmental disturbances}
\label{envir} The torque sensitivity of a torsion pendulum is
intrinsically limited by mechanical thermal noise with power
spectrum
\begin{math}
S_{N_{th}}(\omega)=4 k_B T \Gamma/(\omega Q)
\end{math} \cite{hueller,saulson}
and by the additive noise of the readout
\begin{math}S_{\phi_{read}}(\omega)\end{math}, which can be converted in
an equivalent torque noise by the pendulum transfer function
\begin{math}F(\omega) = [\Gamma (1-(\omega/\omega_0)^2+ i / Q)]^{-1}\end{math} to give
an overall torque sensitivity:
\begin{equation}
    S^{1/2}_{N}(\omega) = \sqrt{S_{N_{th}}(\omega)+\frac{S_{\phi_{read}}(\omega)}{\left\vert
    F(\omega)\right\vert^2}}.
\label{torquetotal}
\end{equation}
Here the torsion pendulum is characterized by the resonance
frequency $\omega_0$, the quality factor $Q$, the moment of
inertia $I$ and the torsional spring constant $\Gamma
=I\omega_0^2$, which for our experiment (see \sref{details}) is
$\approx$ 5 nN m/rad.

\begin{figure}
\begin{center}
  \includegraphics{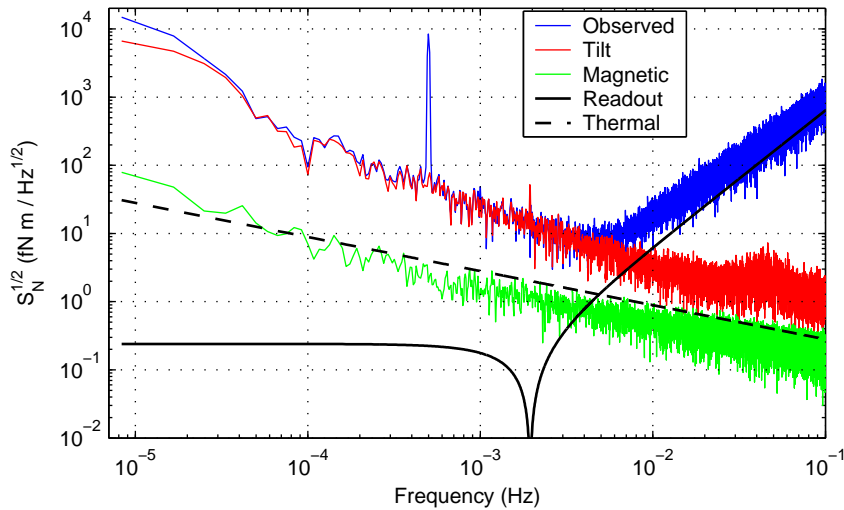}\\
  \caption{Torque noise contributions from external
  environmental couplings, shown with the thermal noise, readout limit,
  and raw torque data. The sharp peak at 0.5 mHz is
  an artifact of a test mass charge measurement \cite{weber:DC} performed during
  the 101 hour run.}
  \label{contributions}
\end{center}
\end{figure}

The pendulum instrumental limit \eref{torquetotal}, in particular
the low frequency thermal noise ($S^{1/2}_{N_{th}} \approx $ 3 fN
m$/\sqrt{\textrm{Hz}}$ at 1 mHz), puts an intrinsic limit on the
resolution with which we can characterize stray forces for LISA.
This can be far exceeded, however, by pendulum coupling to
environmental noise sources. We address four key disturbance
categories:
\begin{itemize}
    \item coupling to the laboratory floor tilt
    \item magnetic field noise
    \item temperature fluctuations
    \item gravity gradient fluctuations.
\end{itemize}
The single test mass configuration, with a compact design and
hollow test mass, makes the gravity gradient noise negligible. We
characterize the other three disturbances by experiments in which
the external source was modulated at a high enough level to induce
a well resolved signal in the pendulum twist. The coupling to each
disturbance term was estimated by the ratio of the induced torque
to the magnitude of the input parameter. Monitoring the
environmental noise levels under normal operation conditions, as
shown in \fref{envirnoise}, permits an estimation of each
systematic effect contribution to the overall torque noise, as
summarized in \fref{contributions}.

\subsection{"Tilt-twist" coupling}
\label{tilttwist} Any tilt motion of laboratory floor can induce a
torque on the pendulum through several mechanisms. To first order,
the torque on the pendulum can be written as
\begin{math} N_T = \frac{\partial N}{\partial \theta} \Delta \theta +
\frac{\partial N}{\partial \eta} \Delta \eta \end{math}, where
$\Delta \theta$ and $\Delta \eta$ are the tilt angles measured by
the sensor itself, as discussed in \sref{details}. The tilt-twist
mechanisms include any position dependent torque induced by the
capacitive sensor, or the known effect of linear cross-coupling of
suspension point tilt into pendulum twist \cite{smith}. A high
immunity from this effect can be gained by making the suspension
point as symmetric as possible and employing the upper pendulum
stage (see \sref{scheme}).

\begin{figure}
\begin{center}
  \includegraphics{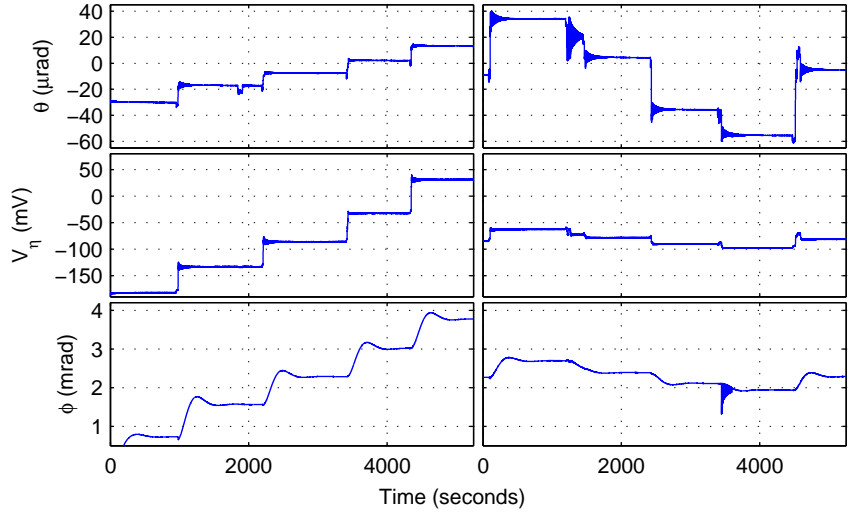}\\
  \caption{Time series of the tilt-twist measurement:
  the experimental platform was tilted along 2 axes separated by 60$^\circ$
  (shown on left and right panels), to induce a tilt and evaluate
  the corresponding change in the fiber equilibrium position.
  The $\eta$-sensor output $V_{\eta}$ was uncalibrated
for rotational sensitivity, but the measured coefficient
$\frac{\partial N}{\partial V_{\eta}}$ is sufficient for
correlating and subtracting tilt-twist noise as described in
\sref{noise}. }
  \label{tilttwistmeas}
\end{center}
\end{figure}

This coupling was measured by purposely tilting the apparatus and
measuring the variation of the rotational equilibrium position
$\Delta \phi$ (see \Fref{tilttwistmeas}). From the DC fiber twist
we can evaluate the induced torque $N_T=F(\omega) \Delta \phi
\approx \Gamma \Delta \phi $ to obtain the coupling coefficients
$\frac{\partial N}{\partial \theta}$ and $\frac{\partial
N}{\partial \eta}$.
The tilt noise along the two horizontal axes was measured by the
sensor, and we can estimate the tilt-induced stray torque to
be\footnote{What we actually measured was the uncalibrated sensor
voltage, $V_{\eta}$, and coupling coefficient $\frac{\partial
N}{\partial V_{\eta}}$. However, the absolute angular calibration
is not necessary, because $ \left\vert \frac{\partial N}{\partial
\eta} \right\vert^2 S_{\eta} = \left\vert \frac{\partial
N}{\partial V_{\eta}} \right\vert^2 S_{V_{\eta}} $.}
\begin{equation}\label{ttnoise}
    S_{N_T} \approx \left\vert \frac{\partial N}{\partial \theta} \right\vert^2 S_{\theta} + \left\vert\frac{\partial N}{\partial \eta}\right\vert^2 S_{\eta}.
\end{equation} This contribution, with observed couplings
\begin{math}\left\vert\frac{\partial N}{\partial \eta}\right\vert\approx
10^{-7}\end{math} Nm/rad and \begin{math}\left\vert\frac{\partial
N}{\partial \theta}\right\vert\approx 10^{-8}\end{math} Nm/rad,
combined with measured tilt noise $ S_{\theta,\eta}^{1/2} \approx
300 $ nrad $/\sqrt{\textrm{Hz}}$ at 1 mHz, is the dominant source
of excess noise as shown in \fref{contributions}.

The measured tilt-twist coupling is considerably higher than those
previously reported \cite{smith}, and strongly directional along
$\eta$. Its origin is likely to be an electrostatic interaction
between the mirror edges (covered by a dielectric coating), and
the surrounding grounded surfaces of the end stoppers that prevent
the test mass from hitting the capacitive sensor walls. An
additional experiment, in which the sensor was translated along
$x$ and $y$ with respect to the pendulum, produced the same
coupling coefficients, confirming that the coupling is dominated
by the relative pendulum-sensor motion, rather than cross-coupling
in the fiber suspension point. In order to suppress this mechanism
in future measurements, we are coating the entire pendulum (in
particular the dielectric mirror) and all the surrounding surfaces
with gold.

\subsection{Coupling to magnetic field}
The components of the residual magnetic moment of the pendulum
$\overrightarrow{m}$ in the horizontal plane couple with the local
magnetic field $\overrightarrow{B}$ to produce a torque
$\overrightarrow{N_B}=\overrightarrow{m} \times
\overrightarrow{B}$; as a consequence, the magnetic field
fluctuations will induce a torque noise of order
\begin{math}S_{N_B}^{1/2} \approx({m_x^2S_{B_y} + m_y^2S_{B_x}})^{1/2}\end{math}.
In order to suppress this contribution, we used nominally
non-magnetic materials for constructing the pendulum, and
surrounded the experiment by a double $\mu$-metal shield. The
horizontal components of the magnetic moment were measured
applying a $B_{\mathrm{pp}}$ = 200 mG sinusoidal magnetic field,
generated by external coils. We then observed the coherent torque
component in phase with the magnetic field. The measured twist
angle was converted into torque to give the magnetic moment
components $m_{x}$ = 120 $\pm$ 5 nAm$^2$ and $m_{y}$ = 40 $\pm$ 2
nAm$^2$, which are possibly due to a 1 cm long, 250 $\mu$m OD
steel tube used in the fiber attachment. The magnetic field noise
measured along the horizontal axes is of order 20 nT
$/\sqrt{\textrm{Hz}}$ at 1 mHz, inducing a torque noise at the
level of several fN $/\sqrt{\textrm{Hz}}$. This stray torque
contribution, compared with the overall noise in
\fref{contributions}, is just below the thermal noise level, and
it could become a limiting factor once we have reduced the
tilt-twist effect. We are thus modifying the fiber attachment
point by employing a Cu tube, which is expected to give a smaller
magnetic moment.

\subsection{Temperature effects}
\label{temper} Noise mechanisms arising from laboratory
temperature fluctuations include modulation of the a) fiber
equilibrium position (a pure ``temperature-twist'' coupling), b)
apparatus tilt due to differential thermal expansion (which
couples to pendulum motion as described in \sref{tilttwist}), and
c) gain of the readout electronics. In order to investigate the
temperature-twist coupling, we modulated the air temperature
inside the thermal enclosure at low frequency (0.5 mHz) while
monitoring the coherent response of the pendulum twist. Referenced
to the fiber tube temperature, we observed a coupling of $\sim$ 1
mrad/K at the modulation frequency. However, the interpretation of
this coupling coefficient is made difficult by the entanglement of
the various thermal effects and the different degree of thermal
filtering associated with each thermometer. An alternative
estimate of the temperature coupling was made by correlating the
low-frequency pendulum twist and temperature data during a normal
run. This analysis also produced ambiguous results that made
impossible the extraction of a set of well-determined coupling
coefficients for the individual thermometers.

\section{Noise measurements}
\label{noise}
\begin{figure}

\begin{center}
  \includegraphics{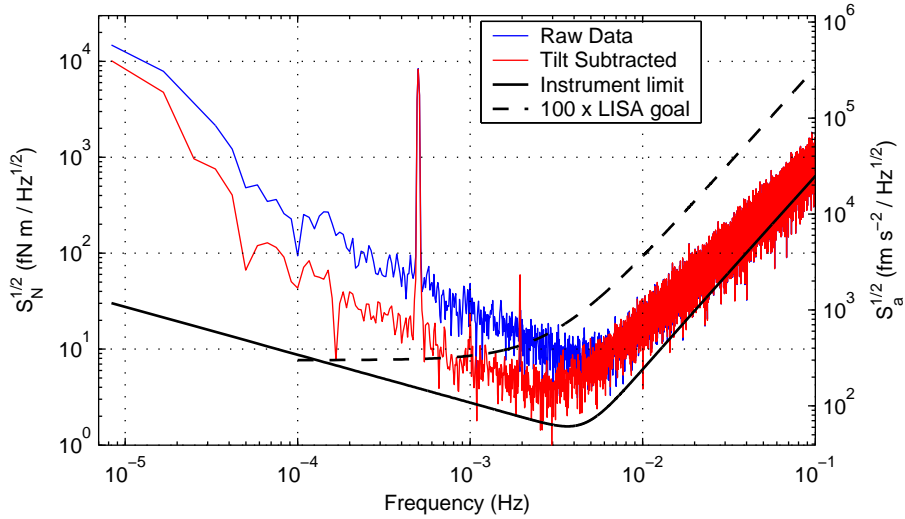}\\
  \caption{Torque and acceleration noise upper limits for LISA,
  calculated from the data shown in \fref{contributions}.
  The acceleration levels, compared with the LISA goal,
  are evaluated assuming an armlenght $a$ = 20 mm to convert torque into force noise,
  and a 1.3 kg cubic test mass (see \sref{LISA}).}
  \label{forcelimits}
\end{center}
\end{figure}

 \Fref{contributions} and \fref{forcelimits} show the typical torque noise level
 $S_{N}^{1/2} = \left\vert F(\omega)\right\vert^{-1} S_{\phi}^{1/2}$,
 and the instrumental limit \eref{torquetotal}.
 The additive readout noise
 dominates at frequencies above 5 $\textrm{mHz}$, where the sensitivity
 of the apparatus is quickly degraded by the $\omega^{-2}$ factor
 in the pendulum transfer function. At $\textrm{mHz}$ frequencies,
 the torque noise is roughly a factor 10
higher than the thermal noise level. As discussed in the previous
section
 and summarized in \fref{contributions}, the excess is
 dominated by the tilt-twist contribution.

 In order to improve our estimate of the low frequency, position-independent
 random forces introduced by the sensor (based on the study of torques on the suspended test mass),
 we subtracted the effect of coupling to floor tilt from the raw experimental data.
 The correction is performed by measuring the components of
 the tilt, and then converting them into a torque by means of the
 measured tilt-twist feedthroughs. The instantaneous coupling torque,
 Fourier transformed into the frequency domain, is then converted into a twist angle
 through the torsion pendulum transfer
 function $F(\omega)$. The calculated twist is converted
 back into the time domain to be subtracted from the raw angular time
 series. This does not
 involve a subtraction of noise spectra, but only a time series subtraction
 based on  calculation of the instantaneous torque. The Fourier transform is
 used only to convert the calculated torques into twist angles, accounting for
the torsion pendulum transfer function. The meaning
 of the subtraction is also checked by comparing the $\theta$
tilt measured with the sensor with the output of the optical
autocollimator, in order to verify that the subtracted signal is a
real apparatus tilt motion, rather than a ``fake'' displacement
signal coming from the sensor itself.

 The tilt correction procedure, whose results are compared with the raw data in \fref{forcelimits},
 leaves a torque noise which is only a factor 3 to 5 over the
 thermal noise in the mHz region, and in particular below 20 fN m $/\sqrt{\textrm{Hz}}$
 between 0.3 and 10 mHz, with a minimum of 4 fN m $/\sqrt{\textrm{Hz}}$ at 3
 mHz.  The excess noise observed
below $\sim$ 0.3 mHz is likely dominated by temperature
fluctuations, although for the reasons described in \sref{temper},
we did not perform any subtraction of temperature effects.

\section{Force noise for LISA}
\label{LISA} The torque noise floor of the pendulum can be
converted into differential force noise, allowing us to place an
upper limit on the stray forces (and accelerations) exerted by the
sensor on the suspended test mass. In order to translate the
results for the pendulum torque noise into stray acceleration for
the envisioned LISA sensors, we should take into account these
constraints:
\begin{itemize}
    \item The torsion pendulum described here was designed to reach
     high sensitivity for surface forces, which are
    expected to be the most dangerous and unknown disturbance
    sources for capacitive LISA gravitational sensors \cite{weber:sens}.
    We employ a hollow test mass, which is largely immune
    to bulk magnetic or gravitational effects, and allows us to increase
    the torque sensitivity with a thinner fiber \cite{hueller}.
    \item The torsion pendulum is highly isolated from net
    forces, generated, for example, by linear field or temperature
    gradients \cite{dolesi}. A future pendulum with multiple test masses, displaced
     from the fiber axis, in order to gain a significant
     conversion between linear (force) effects and measured
     angular twist, will be built.
\item The LISA acceleration levels in \fref{forcelimits} have been
evaluated assuming replacement of the hollow cubic test mass (40
mm
  per side), with a solid Au/Pt mass with the same dimensions,
giving a mass of $m \approx 1.3$ kg. The current design calls for
a 46 mm, 2 kg test mass \cite{dolesi,weber:sens}.
\end{itemize}
The conversion between the estimated torque noise, $S^{1/2}_N$,
and force, $S^{1/2}_f$, for the noise sources under consideration
involves the definition of an effective armlength $a \equiv
S^{1/2}_N/S^{1/2}_f$, which depends on the nature of the source.
For back action forces generated by down conversion of 100 kHz
sensor noise mixing with the readout excitation, or by beating
between stray DC bias and low frequency voltage noise
\cite{weber:sens,weber:DC}, the only non zero contributions to the
instantaneous force $F_x$ on the test mass along $x$ and torque
$N_\phi$ around the fiber axis come from the $x$ electrodes facing
the test mass. Within the infinite wedge approximation we have
$\frac{\partial C_{i}}{\partial x} = \pm \frac{C_0}{d} \neq 0$ and
$\frac{\partial C_{i}}{\partial \phi} = \pm \frac{C_0}{d} R_{\phi}
\neq 0$, where $d$ is the test mass - electrode gap, $C_0$ is the
capacitance formed by each electrode with the centered test mass,
and $R_{\phi}$ (the 1/2 electrodes separation, see \fref{scheme})
is the ratio between the dispacement and angular derivative of
capacitances. Thus, for these effects related to the readout and
actuation circuitry, $a = R_{\phi}$ = 10.25 mm, and the upper
limit can be set at $\approx$ 1.5 pm $\textrm{s}^{-2}
/\sqrt{\textrm{Hz}}$ in the frequency range 0.3 to 10 mHz, when
referred to a bulk LISA test mass of the same size. For forces
generated by any homogeneously distributed forces acting
perpendicularly to the test mass surfaces, such as noisy patch
charges, the relevant armlength is related to the test mass edge
length $s$ by
\begin{math}a = \sqrt{2} \sqrt{\frac{1}{s} \int_{-s/2}^{s/2}
x^2 dx} =\end{math} 16.3 mm, where the integral is an average
square radius, and the factor $\sqrt{2}$ accounts for the
additional torque contributed by the $y$ surfaces. For random
inelastic molecular impacts, the correct conversion is
\begin{math}a = s/2 = 20 \end{math} mm, and the torque noise
levels can be converted into a minimum acceleration noise below
200 fm $\textrm{s}^{-2}/\sqrt{\textrm{Hz}}$ at 3 mHz. As shown in
\fref{forcelimits}, this limit corresponds to roughly a factor 70
over the LISA flight goal, and a factor 7 over the LTP flight test
goal \cite{LTP:1}.

\section{Conclusions}

The experimental campaign we described here was  brought to a
natural conclusion after six months of continuous operation when
the fiber attachment failed. Currently, the apparatus is being
enhanced in order to overcome the sources of coupling to the
environmental noise sources we singled out during the previous
run, including a few functional upgrades. A motorized rotating
stage will allow modulation of the sensor rotation angle $\phi$,
thus characterizing the entire spring-like coupling between the
test mass and the sensor, including any gravitational
contribution. A set of optical fibers carrying UV light will be
used to control the test mass charge \cite{jafry}, and a pattern
of heat exchangers will be used to investigate
thermal-gradient-related effects.

 In addition, we are replacing the gold-coated
torsion fiber with one of bare W, to increase the quality factor,
and thus decrease the thermal noise. In combination with the
increased immunity from environmental systematic effects, we will
be able to characterize LISA sensors with higher sensitivity.

\ack It is a pleasure to acknowledge many fruitful discussions
with E Adelberger.

\end{document}